\documentclass[letterpaper,twocolumn,10pt]{article}
\usepackage[utf8]{inputenc}
\usepackage{usenix,epsfig}
\usepackage{threeparttable}
\usepackage{hyperref}
\usepackage{xspace}
\usepackage{tabu}
\usepackage{balance}
\usepackage{color}

\newif\ifDraft
\Drafttrue  

\ifDraft

\else

\fi

\newcommand{\vendorkeymgr}{Manufacturer Key Manager\xspace}

\newcommand{\guardian}{\textsf{Guardian}\xspace}
\newcommand{\key}{\mathit{K}}
\newcommand{\device}{\mathit{d}}
\newcommand{\guard}{\mathit{G}}
\newcommand{\update}{\mathit{sw}}
\newcommand{\pw}{\mathit{pw}}
\newcommand{\AP}{\mathit{AP}}		
\newcommand{\trustdom}{\mathit{T}} 	

\begin{document}

\date{}

\title{Baseline Functionality for Security and Control of Commodity IoT Devices\\ 
 and Domain-controlled 
 Device Lifecycle Management
} 

\author{
{\rm Markus Miettinen$\dagger$}\\
\and
{\rm Paul C. van Oorschot$\ddagger$}\\ 
\and
{\rm Ahmad-Reza Sadeghi$\dagger$}\\
\and
{$\dagger$TU Darmstadt, Germany   $~~~\ddagger$Carleton Univ., Canada} 
}


\maketitle

\thispagestyle{empty}

\subsection*{Abstract}
The emerging Internet of Things (IoT) drastically increases the number of connected devices in homes, workplaces and smart city infrastructures. 
This drives a need for means to not only ensure confidentiality of device-related communications,
but for device configuration and management---ensuring that only legitimate devices are 
granted privileges 
to a local domain, 
that only authorized agents 
have access to the device 
and data it holds, 
and that software updates are authentic. 
The need to support device on-boarding, ongoing device management and control,
and secure decommissioning
dictates a suite of key management services
for both access control to devices, and access by devices to wireless infrastructure and networked resources. 
We identify this core functionality, and argue for the recognition of
efficient and reliable key management support---both within IoT devices,
and by a unifying external management platform---as 
a baseline requirement for an IoT world.  
We present a framework architecture to facilitate secure, flexible and convenient device management 
in commodity IoT scenarios, and offer an illustrative set of protocols as a base solution---not
to promote specific solution details, 
but to highlight baseline functionality to help domain owners
oversee deployments of large numbers of independent 
multi-vendor IoT devices.



\section{Introduction}   \label{sect:intro}

The Internet of Things (IoT) involves connecting and managing enormous numbers of  devices   
in wide-ranging applications, from commodity retail products to 
advanced medical devices 
and autonomous vehicles. 
The scale is such that traditional solutions for device association and management become obsolete. The problem is exacerbated by individual IoT device manufacturers promoting vendor-specific solutions for device onboarding and management; unsurprisingly, interoperable key management solutions for IoT devices are as yet unknown.

This increases the burden on the owner of an 
IoT \emph{trust domain}---e.g., an IoT-enabled residence, 
industrial building, 
or metropolitan 
infrastructure---as numerous distinct IoT device
management applications must be used for common tasks
related to connectivity and security such as
software/firmware updates, 
and changing passwords to a domain wireless access point.
Many device manufacturers provide only rudimentary management tools, 
leaving keying, configuration and software management for domain owners
to address manually, or removing all user control over when or if
software updates occur. Beyond major new burdens, possible mis-configuration puts at risk the security of entire trust domains.

To address these issues, we outline a novel key managment framework for
commodity IoT domains. It adopts a \emph{guardian} model \cite{Barrera:2011:SPmag}.
A configurable intermediate entity, the KMI (Key Management Infrastructure)  \guardian{}, controls security-relevant aspects of the lifecycle of IoT devices. We ground the framework architecture on a taxonomy of IoT devices (Section~\ref{sect:iot-environments}), which divides IoT devices into three broad qualitative categories (\emph{high-end}, \emph{mid-level}, \emph{low-end}) based on computational and communication capabilities. As we assume premium 
devices will follow advanced 
vendor-specific security frameworks, 
our framework focuses on commodity 
devices, providing support for authentication, authorization and integrity verification.

The \guardian{} manager is designed to work in parallel with
vendor-specific, device-specific applications which control 
fine-grained IoT device functionality; it aggregates lifecycle aspects related to  
access control, network connectivity, security and privacy,
addressing head-on the complexity arising from 
ownership of large numbers of heterogeneous devices from a wide
spectrum of international manufacturers with whom domain owners have no
meaningful relationship, yet whose commodity devices require ongoing, individual 
care and feeding.


\section{IoT Landscape (Use Environments)}  \label{sect:iot-environments}

The space of IoT devices is booming, with an exploding number of device manufacturers  introducing new devices with network connectivity. The related promise of the ability to supervise and control devices remotely has led the adoption of an IoT paradigm in a variety of application scenarios. We mention three briefly:  
smart homes, smart cities/buildings, and industrial IoT.
We then briefly discuss categories of IoT devices.

\textbf{1) Smart Homes.}
IoT in smart homes currently targets small-scale, commodity devices allowing  homeowners to monitor, control and automate elements of their home environment. Typical devices in this setting include smart light bulbs, power plugs, doorbells, IP cameras, and a multitude of sensor-enabled appliances. 
Sophistication level and price range varies greatly. 
Countless device vendors already offer products in this category, with significant heterogeneity in device design and software solutions.
Among premium manufacturers of smart home devices 
(e.g., Samsung, Apple, Google/Nest), some devices are part of
full-stack product solutions with 
integrated device-family management and operation. More commonly, management solutions lack integration, with independent means for on-boarding, access control and managing software updates; 
we refer to these as \textit{commodity devices}.

\textbf{2) Smart Cities and Buildings.}
The building of IoT devices into infrastructures is referred to as \emph{smart city} or \emph{smart building} scenarios. These may involve devices that satisfy special requirements, e.g., with regard to protection against rough environmental conditions caused by rain, dust, heat or low temperatures. Devices in these scenarios may also employ special long-range radio communication protocols or cellular access instead of `standard' WiFi or Ethernet-based communication links, in order to be deployable in a wide area with different environmental settings.
These specialized scenarios with highly customized devices in large-scale device networks with comparatively high per-unit costs mandate device vendors to provide customized solutions for device deployment and management 
allowing automated, centralized mass provisioning of devices. These fall outside our key management architecture for commodity devices.

\textbf{3) Industrial IoT.}
As with smart city/smart building settings, industrial solutions are characterized by significantly larger scale of the targeted systems and special environmental requirements towards devices. Also here, management solutions are tailored by system vendors to meet special requirements of the deployment environment of particular industrial environments.  
As such, large parts of industrial IoT are beyond the scope of our IoT key management framework, however our framework remains suitable for smaller-scale industrial environments, and individual departments within larger organizations, when these units use commodity IoT devices.

\textbf {Categories of IoT devices (taxonomy)}.
IoT devices can be characterized according to their properties related to: processing capabilities, memory, operating system, energy consumption, communication protocols and encryption capabilities. For background context, Appendix \ref{sect:properties} discusses such characteristics, with focus on IoT devices typical in smart home environments, the main target for our proposal. Appendix \ref{sect:low-med-high} then uses these to classify IoT devices in three categories: \emph{high-end}, \emph{mid-level} and \emph{low-end}.  
Smart home devices in our target scope fall  
in the mid-level and low-end categories, which
have sufficient computing and memory resources to execute 
suitably customized (e.g., ECDH PAKE) protocols for establishing keying relationships with the \guardian{} per Section  \ref{sec:guard-onboard}.
Beyond our scope are devices---not always 
more powerful in the sense of being high-end (above)---from 
premium vendors 
providing vendor-specific on-boarding and device-family
management applications integrated with on-line services.
	
At the lowest-capability end of the commodity consumer device market
are ``extremely constrained" devices, 
not prevalent in smart home networks. 
Terminology for constrained, IP-connected nodes is defined by
RFC 7228\footnote{\url{https://tools.ietf.org/html/rfc7228}}, distinguishing  
Class $0$, $1$ and $2$ nodes based on resources
available on low-end devices.  The most constrained (Class 0) devices fall
clearly below our scope---most
simple sensor devices we have examined provide significantly more resources. 
All devices we classify as low-end (Table~\ref{tab:iot-device-details}, Appendix~\ref{sect:low-med-high}) fall in RFC 7228 Classes 1 and 2; 
for context, our mid-level devices are clearly beyond Class 2 capabilities.




\section{Key Management supporting Life-Cycle}   \label{sect:requirements}

An IoT key management system must support operations involving crypto
keys throughout the typical life cycle of IoT devices. As discussed here, these include:
device onboarding (establishing an initial keying relationship with a new device); 
key lifecycle management; 
administrative access control to devices; 
verification of software updates; 
and
integrity verification of device state.

IoT key management requirements follow closely from the life-cycle of
an IoT device. Typical stages and relevant events are as follows. 
\begin{enumerate}

\item [S1:]
Device registration and \textit{on-boarding}. Installation in trust domain $\trustdom$ includes: 
establishing connectivity with $\trustdom$, establishing keys for authentication, access control,
and secure communication. 

\item [S2:]
Configuration and mainstream use. This includes:
secure transmission of device sensor data; 
secure remote access by owner (to examine state, change parameters);
event logging if appropriate. 

\item [S3:]
Software/firmware update.  For simplicity, we assume that essential 
state is maintained across any power losses or battery replacements.

\item [S4:]
Device removal from domain (decommissioning,
e.g., due to obsolesence or transfer to new owner).
This includes: deleting sensitive state,
including recorded data, domain passwords and keys. 

\end{enumerate}
We next comment on aspects of these stages.

\textbf{Device on-boarding.}
In typical IoT scenarios, no prior security context exists between a \emph{trust domain} and a device to be added to it. Keying relationships must thus be set up. Traditional ad-hoc key establishment means face challenges in IoT scenarios---e.g., the number of devices precludes extensive user involvement in terms of physical contact with the device, or of entering or verifying authentication secrets between the onboarded device and another domain-related management device.

Solutions requiring physical contact with each device may be impractical in terms of manual labor cost in IoT scenarios involving tens or hundreds of devices (home environments), not to mention thousands of devices 
(e.g., smart bulbs in industrial IoT or smart building scenarios). Such physical electrical contact with a new device for securely transferring key material over this connection have been proposed, e.g., in the \emph{resurrecting duckling} model~\cite{Stajano:1999:workshop}.    
Approaches like \emph{Bluetooth Secure Simple Pairing}\footnote{\url{https://www.bluetooth.com/specifications}} and \emph{WiFi Protected Setup}\footnote{\url{https://www.wi-fi.org/discover-wi-fi/wi-fi-protected-setup}} use common passwords for authenticated key establishment between devices. However, manual effort required limits scalability, and many simple devices lack a suitable user interface to input or verify passwords or authentication secrets. Repeated manual entry or verification of such secrets is also error-prone, putting at risk the security of the approach. 

What is needed is a (retail-level) scalable approach for establishing keying relationships for large numbers of devices, securely and reliably, avoiding anything beyond very minor
per-device human involvement or oversight.  

\textbf{Life-cycle key management.}
IoT devices may store information and access credentials to other systems (e.g., the vendor cloud) that enable interaction with other devices in the trust domain and access to common (configuration) information of the domain. This access should depend on device membership in the domain. On any change in domain membership, access to domain resources should be updated accordingly. In particular, if membership is revoked or expires (or the device is transferred to a new owner), all existing domain-related access credentials should be revoked.
Effective means are thus needed for managing device-held credentials for accessing other devices and resources both within and outside the local trust domain.

\textbf{Device access control.}
Some IoT devices provide access to APIs or direct terminal access for device administration and control. IoT malware (e.g., \emph{Mirai}, \emph{Hajime}) have utilized weak or default passwords on administrative access interfaces (e.g., \texttt{ssh}, \texttt{telnet}) to gain unauthorized device access and install malicious code.
It is thus necessary to protect access vectors through suitable authentication keys 
(e.g., not easily guessed by 
IoT malware), and to ensure that any default passwords used are replaced or disabled on domain admission.

\textbf{Verifying software/firmware updates.}
Software flaws, new features and optimizations result in post-sales software (and firmware) updates; moreover, some
IoT devices may last 10-20 years, by which time vendors may be out of business. 
To protect devices from unauthorized or flawed updates, update integrity and authenticity must be verified. In some cases, update is irreversible and updating to a dysfunctional image may render a device permanently unusable. 
A trust domain owner should also be able to control which software updates are installed on each device, 
\emph{independently} of the device vendor or other update source.




\section{Current Practice} \label{sect:threat-model}
 
Our framework is motivated by issues in current practice.

\begin{figure}
	\includegraphics[width=\columnwidth]{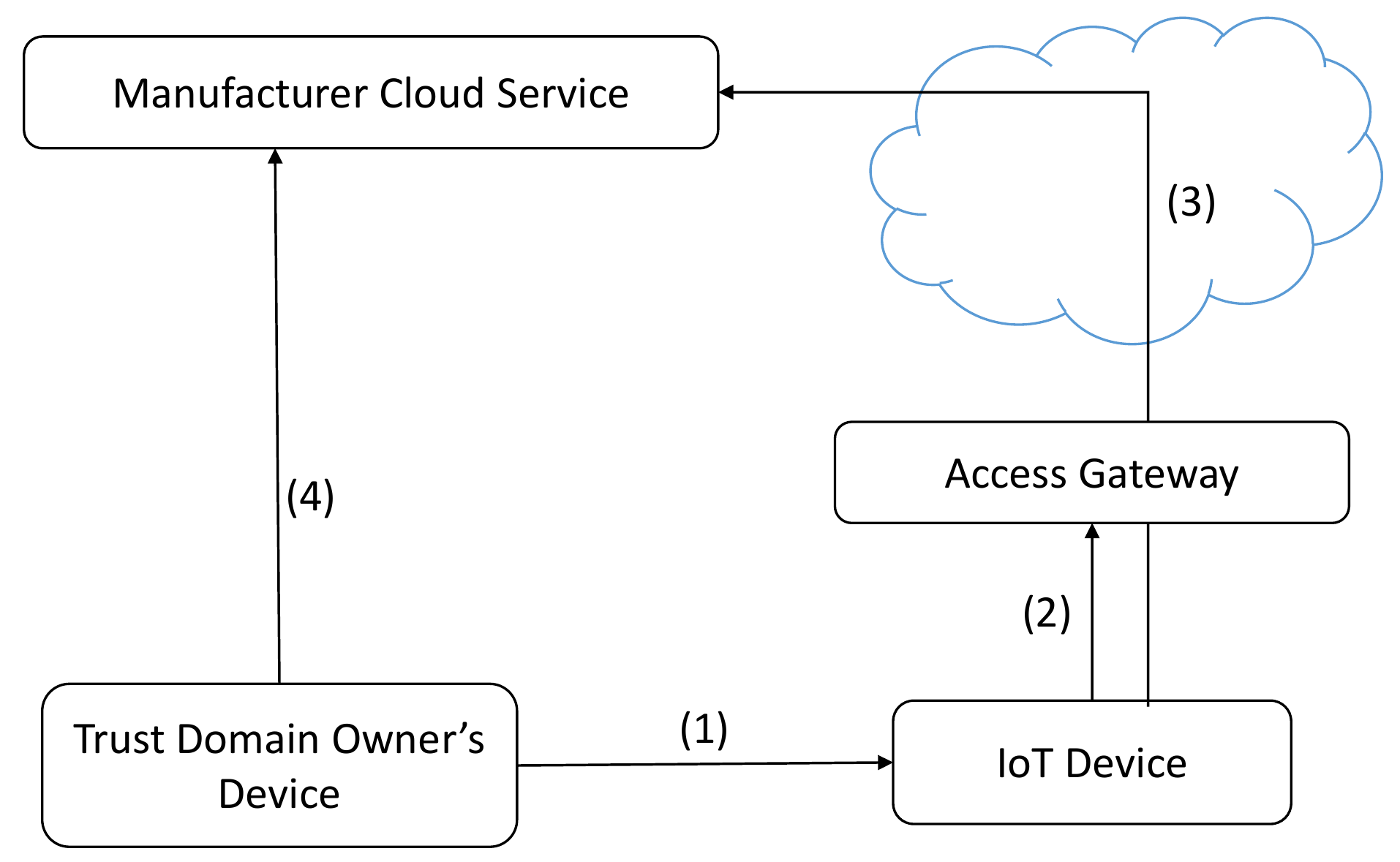}
	\caption{Typical process of establishing a keying relationship with current IoT devices.}
	\label{fig:soa-pairing}
\end{figure}

\textbf{Current practice (device on-boarding).}
Current IoT devices seldom use physical contact for authenticating a keying relationship, in part due to the lack of standardized physical interfaces and protocols for key distribution over physical connections. Keying establishment involving WiFi IoT devices more commonly proceed as follows (Figure \ref{fig:soa-pairing}). The domain owner installs a smartphone companion app from the device vendor, often requiring the domain owner to register with a device vendor's on-line service for enhanced functionality (e.g., remote access to device state while the owner is beyond the device's local network). On initial power-up, the device sets up an ad-hoc WiFi access point for the domain owner to connect to by smartphone companion app, transferring credentials (1) for IoT device access to the trust domain's WiFi network. The device then 
connects to the local access gateway, using the credentials received (2). After establishing network connectivity through this gateway, the device may connect to a vendor cloud service (3) for any software updates, and may provide remote access to its device state. In parallel, domain owner's smartphone connects to this cloud service (4) and registers the IoT device, tying it to an owner user account.

This widely-used onboarding approach has deficiencies limiting its usefulness and security. 
Typically during on-boarding, the domain owner must manually identify a target IoT device's WiFi access point (AP) to connect to it (some companion apps attempt automated choices). 
However, it is difficult to distinguish the intended device's AP (SSIDs are often obscure), 
from a rogue AP, or a
non-malicious AP of a nearby unprovisioned device not within the trust domain.
Device manufacturers may use vendor-specific device certificates to verify AP authenticity to combat spoofing attacks, but this does not preclude
the app connecting to a nearby (within wireless range) genuine same-vendor device that is not part of the trust domain $\trustdom$.

\textbf{Current practice (other aspects).}
Current practice generally provides no guidance or accommodation
for stages described below (in our model) as
generating and keeping a device inventory,
mediating software updates, 
or decommissioning devices.

\begin{figure}
	\includegraphics[width=\columnwidth]{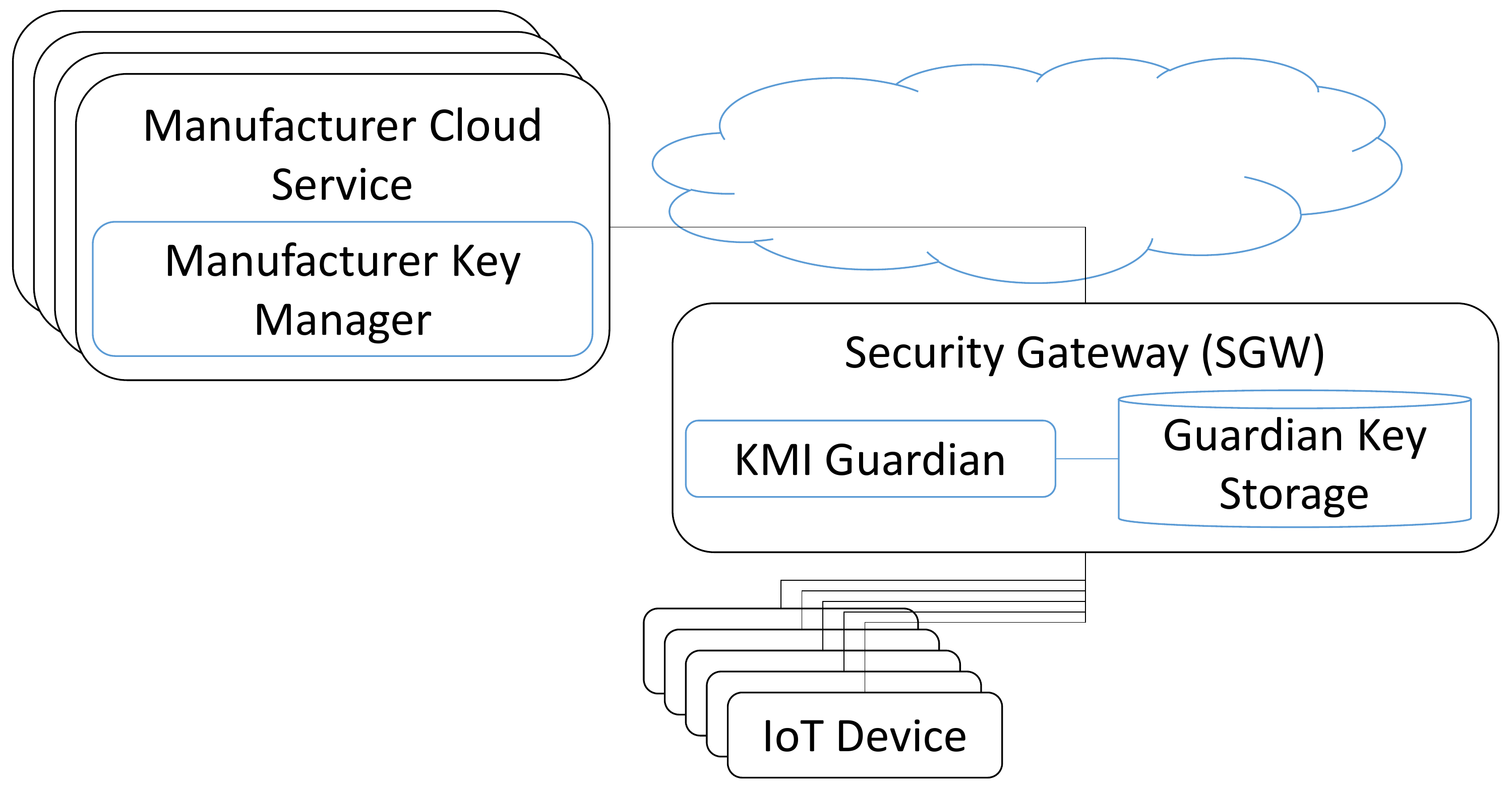}
	\caption{Guardian architecture for IoT key management infrastructure and security gateway.}
	\label{fig:system-model}
\end{figure}

\section{Architecture and System Model}  \label{subsec:sysmodel}
We envisage (Fig.~\ref{fig:system-model})  IoT devices belonging to a  trust domain managed by a component, the \guardian{}, located on a local Security Gateway (SGW), taking the role of typical home WiFi routers. 
The \guardian manages all device keys $\key$ issued to IoT devices $\device$ in the local trust domain. It provides a management console and support for on-boarding,
key establishment (and updating/revocation), and both verification and local approval of software updates for individual devices $\device$.
For certain devices, the \guardian may co-operate with the \vendorkeymgr of an IoT device to establish the initial keying relationship.


\textbf{Threat Model.} \label{subsec:threatmodel}
We assume that attackers begin 
without legitimate authentication credentials/keys.
We divide adversaries into two classes: those with
only remote access, and those with local access.
Remote attackers have (wireline) access to Internet-addressable
devices.  Local attackers have access to one or more of:
the local wireline network (LAN);
wireless interfaces (of individual IoT devices and the Guardian);
and physical access to IoT devices.
We assume that the Security Gateway with the \guardian is physically
protected by other means, and do not further   
consider targeted attacks that physically access it.
Threats we discuss are as follows. 
\begin{enumerate} 

\item [T1:]
Remote attackers trying control or manipulate IoT devices through forged software updates.

\item [T2:]
Remote attackers seeking to bypass access control in order to
access local state or reconfigure devices.  

\item [T3:]
Remote attackers aiming, other than by the update process,
to infect IoT devices with malware,
including the goal of denial-of-service attacks.

\item [T4:]
Local attackers within range of IoT device wireless interfaces,
with any of the above objectives.
For example, this includes 
a rogue companion app connecting to a legitimate IoT device AP; and 
a rogue IoT device being admitted to the trust domain.
The attacker may aim to establish a keying relationship by
tricking either device $\device$, or the $\guardian$, in connecting to its own
(rogue) device.

%
 
\item [T5:]
Local attackers with physical access to a domain's IoT devices
are beyond our scope 
of commodity IoT devices and non-targeted, scalable attacks.
Physical attacks require more expensive defenses
(see TEEs, Section \ref{sect:background}).
We argue that
before attempting to find cost-efficient solutions to targeted attacks,
the community is best served by entrenching baseline solutions
to simpler problems unaddressed in current practice.
We comment further on complementary use of TEEs in the concluding remarks.
%
%

\end{enumerate}

Attacks should be expected on the protocol
between the \guardian and a device $\device$ during on-boarding,  
to reconstruct shared secrets;
thus a secure protocol is necessary (PAKE, below). 
A general attack sub-goal is to 
gain access credentials to the trust domain $\trustdom$.




\section{Guardian Design: Functionality}  \label{sect:arch}

We give illustrative solutions to clarify the baseline functionality
and concrete issues it addresses;
the specific solution details are not the focus \textit{per se},  
albeit providing an instructive base from which to build.
We assume (assumption A1) that all in-scope IoT devices can be ``hard reset" 
putting the device in   
a provisioning state ready to establish a device communications key $\key$ ($\key_{G,d}$ below);
many existing IoT devices have such a physical button.
 
\textbf{Notation.} 
$\device$ (IoT device), 
$\guard$ (Guardian unit),
$\trustdom$ (trust domain), 
$\trustdom_\AP$ and $\device_{\AP}$
(WiFi access points of $\trustdom$ and $\device$).

\subsection{Guardian device registry (inventory)    \label{sec:guardian-rollcall}}

The Guardian is responsible for keeping an \textit{inventory}
(\textit{device registry})
of IoT devices authorized to be part of the trust domain. Each device
entry contains:
\begin{itemize}
\item
unique device ID (e.g., unique serial number)
\item
hardware MAC address (as a secondary ID)
\item
device manufacturer and unit model
\item
common-language description (``smart plug", ``lightbulb type xxx",
``door-lock", ``IP camera") 
\item
software/firmware version currently installed  
\item
software updates available (manufacturer; other),
with reason and/or urgency code motivating each (stability, security, functionality)
\item
(protected) keying material, including device-specific access password to 
$\trustdom_\AP$, the Guardian or trust domain AP.

\end{itemize}   

Note: device-specific passwords to $\trustdom_\AP$ address a common flaw: 
when devices use the same password to access a home
WiFi AP, password compromise puts all devices at risk, 
yet the inconvenience of reconfiguring them all discourages changing the password.
$\trustdom_\AP$ 
looks not for a fixed password, but one based on
incoming device MAC addresses (from link level/Layer 2), which
indexes a table of device-specific verification data (cf.\ Unix).

Prior to on-boarding (Section \ref{sec:guard-onboard}), 
which populates additional registry fields,
a rostering process creates device entries as follows. 
Assume IoT devices have a physical sticker with QR-code (2D-barcode),
indicating a serial number and device-specific password for the device's own
wireless interface.  
The sticker is scanned using a companion app, creating 
inventory entries for all devices to be on-boarded, populated with the
device ID and this to-device password; information such as 
MAC address and currently installed software version can be automatically
acquired in later device interactions, while other fields may be
auto-populated including from online sources to
reduce the information needed on QR-codes,
e.g., serial number may imply manufacturer, model, description.
While requiring per-device scanning, 
this process avoids manual creation of device entries. 

Any device not physically scanned (thus not in the inventory) is 
precluded from on-boarding.
An attacker aiming to admit rogue devices into a trust domain must introduce 
such devices into the pool scanned by the trust domain owner. 
While a low barrier to inside-attackers in 
industrial scenarios, we argue this provides a good
cost-benefit tradeoff in the target commodity 
IoT scenario.

In current practice (Section \ref{sect:threat-model}),
default on-boarding passwords are common. For devices of reasonable size,
vendors afix stickers with device-specific random passwords,
which may be derived deterministically from scanable device IDs 
(serial number or MAC address) plus optionally a vendor secret; 
an assembly-line test/packaging process scans the device, computes the password,
afixes the sticker.  
The to-device password is at the same time provisioned into the device, e.g., by flashing to a particular persistent storage memory address.
The value of such passwords erodes seriously if they are global defaults  
(known to buyers and attackers alike), but may still aid on-boarding 
absent active attackers; but device-specific to-device passwords are
already common in commodity IoT devices.
 

\subsection{Guardian-based on-boarding}   \label{sec:guard-onboard}

\begin{figure*}
	\includegraphics[width=\textwidth]{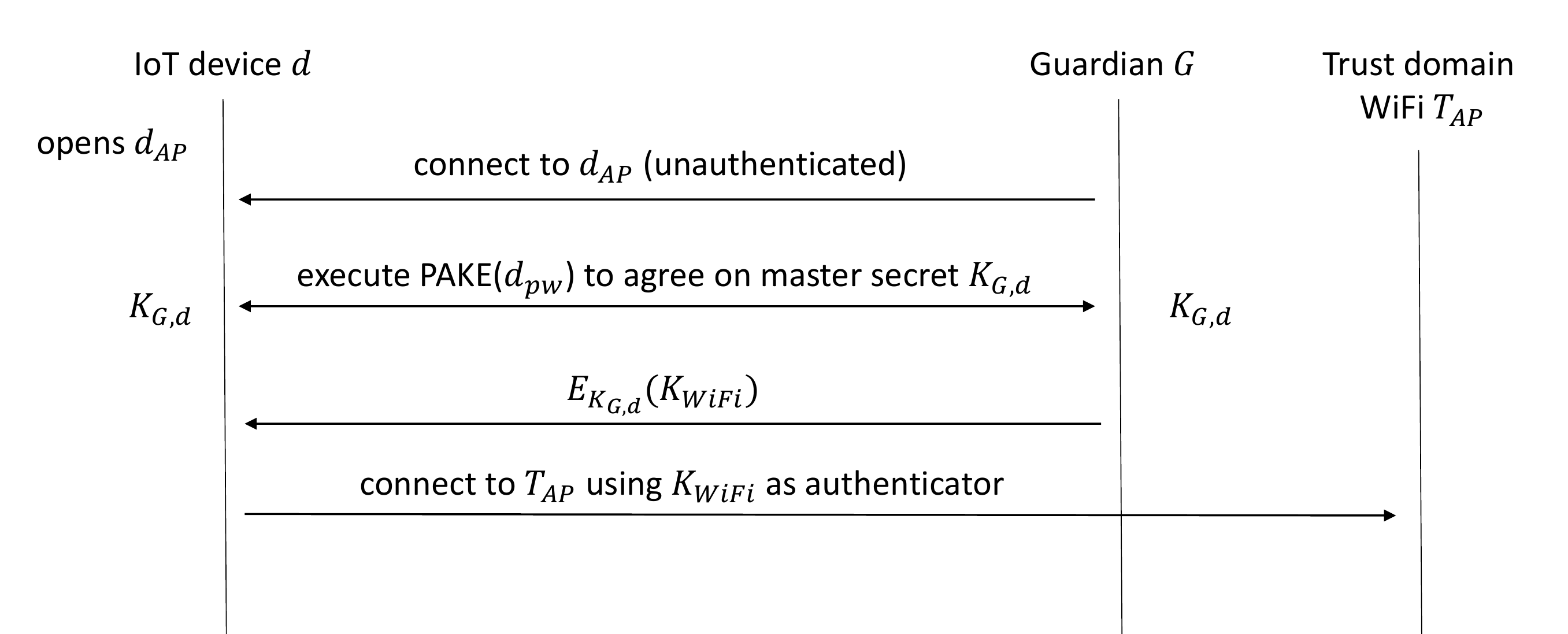}
	\caption{Guardian-based on-boarding of an IoT device}
	\label{fig:on-boarding}
\end{figure*}

We first outline an illustrative on-boarding process for concreteness. 
It uses a password-authenticated key exchange (PAKE) protocol,\footnote{We
suggest J-PAKE \cite{hao:2008:protocolwksp,abdalla:2015:oakland} for practical reasons.} 
default-username, and password $\device_{\pw}$ pre-configured into $\device$, as shown in Figure \ref{fig:on-boarding}.
This key establishment process delivers device-specific \textit{crypto-quality working keys}.

\begin{enumerate}
	\item Bring $\device$ into configuration mode (hard reset)
	\item $\device$ sets up its access point $\device_{\AP}$ (open/unprotected) 
\item $\guard$ connects to $\device_{\AP}$ (unauthenticated/open)
\item $\guard$ and $\device$ execute a PAKE using $\device_\pw$ as authenticator to establish a master shared secret $\key_{G,d}$
\item $\guard$ sends $\device$ the WiFi key, encrypted: $E_{\key_{G,d}}(\key_\mathit{WiFi})$ 
\item $\device$ connects to $\trustdom_\AP$ over WPA2 with password $\key_\mathit{WiFi}$
\end{enumerate}
\noindent
As discussed, ideally $\key_\mathit{WiFi}$
is a per-device crypto-grade secret (rather than a common password shared by all
devices using the access point).

An advantage of the above design is that impersonation by an adversary requires the password 
$\device_{\pw}$. 
As noted in Section \ref{sect:iot-environments},
computing requirements imposed by (e.g., customized, ECDH) PAKE
exclude only extremely-constrained devices already outside our scope; also resources  
for public-key operations, which are needed only during device on-boarding (by PAKE),
are increasingly penetrating highly-constrained devices \cite{Ozmen2017}. We argue
that in the near future, almost all consumer IoT devices will have such capabilities,
especially if supported by standards \cite{keoh:2014:journal}.  
Since an active adversary knowing $\device_{\pw}$ could extract $\key_\mathit{WiFi}$, note that
$\device_{\pw}$ must be resistant to trivial online guessing.

To succeed, an attacker knowing $d_{pw}$ must authenticate
before $\guard$ does.  Designing $\device$ to have a 
``has been on-boarded" flag allows this to be detected.

\textbf{Reference monitor option}.
The above on-boarding process gives $\device$ connectivity to $\trustdom_\AP$.   
Alternatively, $\guard$ could mediate all communications to $\trustdom_\AP$ to provide firewall-like features (e.g., 
setting up a circuit-level proxy). 
More generally---though not a focus herein---the
Guardian architecture provides a strong control point and
logical place for other security gateway services including intrusion detection 
and per-device traffic (volume) logging; and
rather than facilitate access to a separate trust domain WiFi AP
(by providing access passwords to it), could
itself be the WiFi AP (or Internet gateway) for consumer IoT devices.
Applying such a network-security based approach to consumer IoT devices, as also suggested by others  
\cite{Simpson2017,Miettinen2017,Barrera2017},
offers strong control and flexibility in mitigating compromised devices.

\subsection{Guardian-mediated software update} \label{sec:guadian-update}

Based on a list of the domain's IoT device models, derived from the registry,  
the Guardian regularly refreshes (based on searches of vendor web sites
and other databases) its list of available
software updates for its device roster, plus  
update reasons (to inform decisions about which updates to allow). 
This not only allows visibility and oversight of  
software and firmware update of domain IoT devices, but
reduces hardware support needed in end-devices
and simplifies software complexity. The simplification stems
from no longer needing end-device cryptographic support for public key    
signature and certificate verification; 
revocation checking (which requires trusted timeclocks); 
and provisioning and renewal of PKI trust anchors
(a complex issue for IoT devices, whose 5-20 year lifetimes  
are guaranteed to span manufacturer reorganizations and bankruptcies).
 
Using its list of IoT devices and available updates, the \guardian{}
decides which updates to allow (based on internal policies and/or external advice).
For each, it 
downloads the update, and verifies its authenticity and integrity after having
aquired and verified any necessary trust anchors. Note this is done in a
single, authoritative, dedicated security platform (the  \guardian{}), rather than
in a myriad of IoT devices for which security is a non-core competency.
For each IoT device in the domain approved to receive this specific
update, the  \guardian{} does the following:  
 \begin{enumerate}
 \item
 puts the device into a state ready to receive 
(from the  \guardian{}) the pre-verified update;

 \item
 transfers the update, secured by a working key (e.g., previously established MAC key);
 
 \item
 instructs the device to accept the update, if the MAC verifies
 (in place of a manufacturer signature).
 
 \end{enumerate}
 \indent
The new baseline functionality of IoT devices must support these actions. The
 cryptographic requirement reduces to low-cost MAC verification.
Trust is based on the pre-establishing crypto keying relationship between device and
\guardian{}.
For 
commodity IoT devices, this can significantly reduce both
processing power required and logistical complexity otherwise needed 
for secure software/firmware update.

\subsection{Guardian de-commissioning of devices} 

The design of the Guardian, and compliant devices, allows the following,
before discarding a specified IoT device for recycling.
\begin{enumerate}
\item
A Guardian-triggered software reset  
achieves the equivalent of a hard device reset:
removing on-device sensitive user data and state, 
including crypto keying material, and
passwords to resources including local WiFi and internal or external resources.

\item

The Guardian, as appropriate, deletes its own copies of all passwords related
to the device, 
keying material, and user-related sensitive data.

\item
If the device is to be re-sold or transferred, a new device initial password
will be needed by the new owner, to access its device AP for on-boarding.
We assume that any password-bearing sticker has been lost; a
baseline solution is outlined next.
\end{enumerate}
\noindent 
The new owner retrieves a serial number 
stamped on the device exterior (in this baseline solution, this
requires visual access to scan the device portion bearing the serial number). 
From that, a default password is retrieved or regenerated, 
with co-operation of a vendor site, in a variation of the means used to
deterministically generate
initial device-instance specific passwords (Section \ref{sec:guardian-rollcall}). 
A central database run by the device vendor or proxy thereof, keeps a
reset count $c$ (indexed by serial number $s$), and knows the corresponding MACaddr. 
A device-specific ``next-password" is then generated by 
combining a vendor secret with (MACaddr, $s$, $c$). 
Requests made to the vendor for a new password, accompanied by $s$, 
result in the vendor updating $c$, logging an asserted owner ID
(or phone number or IP address), and returning the next password in the sequence.  

This returns a password to a new owner,
but that secret would not itself (yet) be in the device---so
the process requires a small twist. This new password 
could be loaded into the device by
a design using a physical interface (thus also ensuring 
physical possession), but this would be user-unfriendly
and have other drawbacks (e.g., the secret could be made up independent of
the vendor, an on-boarding of second-hand devices with self-loaded
secrets may abet device theft).
The suggested process is instead to use standard
Lamport hash chains, i.e., one-time password sequences based on iterated hashing
$h(h(h(...h(w)...)))$ with, say, $t = 200$ steps, for some device-specific
initial secret $w$. Here $w_0 = h^{t}(w)$, the $i^{th}$ password is 
$w_i = h^{t-i}(w)$ and $h(w_i) = w_{i-1}$, so the verifier receiving a password
${w_i}^*$, asserted to be the $i^{th}$ password, tests it by hashing it once 
and checking against the previously stored password $w_{i-1}$, i.e., testing:
does $h({w_i}^*) = w_{i-1}$?
So the device can
easily verify the next on-boarding password (used for exactly one on-boarding),
and $i = c$. There is no security risk in the device making the count $c$ 
(advanced by physically pressing the reset button, or equivalent by a soft-press)
available electronically, without device authentication;
the new owner (new owner's Guardian) can thus obtain it over a wireless interface, 
and send to the vendor site the triplet
(MACaddr, $s$, $c$), to obtain the next on-boarding password.%




\subsection{Summary of baseline functionality}  \label{subsec:goals}

The lifecycle stages outlined in Section \ref{sect:requirements}
map closely to the design aspects outlined in the subsections above.
%
%
We now review security functionality provided.

The \guardian controls and manages
access to individual IoT devices through established crypto-grade keys. This is important as high-profile attacks (e.g., the Mirai botnet) have used unauthorized access to IoT device management interfaces to gain control over victim devices.  
%
The \guardian also has full control over software updates installed on particular IoT devices,
and verifies the updates on behalf of devices. This reduces the risk of remote attackers  installing rogue software. The \guardian mediates which updates are installed,     \textit{independent} of authorization by the device vendor, screening updates based on relevant factors including evaluation of the security, reliability, or reputation of specific updates. 
It keeps a history, for all IoT devices in the domain, of software/firmware versions installed.



We summarize 
baseline functionality provided by the \guardian and related device-end support:
\begin{description}
	\item[B1: Device inventory (rostering).] A registry of all IoT devices
	in the trust domain. Fields include device ID information and
	device-specific password access and crypto-grade keying material. 
	\item[B2: Master key establishment.] Means to establish a master shared secret (device key $\key =\key_{G,d}$) between a new device $\device$ and the trust domain $\trustdom$. From this, device-specific crypto
	working keys are derived, including for encryption and integrity (authenticated
	encryption) of data transfers, and to convey to devices the integrity of software updates approved
	by the \guardian{} $\guard$.
	\item[B3: To-device, from-device authentication.] Means for   
	$\guard$ to authenticate into device $\device$ (i.e., into $\device_{\AP}$,
	initially by a
	manufacturer-provided password; updated by a crypto-grade secret after
	master keying). A derived working key is used for authentication
	of $\device$ to  $\guard$, and a device-specific key for $\device$ to use
	the domain access point $\trustdom_\AP$. 
	\item[B4: Key update management.] Means to update and expire any key $\key_i$ (and all keys derived from it) associated with a specified device $\device$.
	\item[B5: Upgrade discovery.] Means to discover available software/firmware updates
	for rostered IoT devices.
	\item[B6: Update verification.] Means for $\guard$ (rather than end-devices) to verify the integrity and authenticity of software updates $\update$. 
	\item[B7: Upgrade mediation.] Means for $\guard$ to mediate installation of 
	any software update $\update$ to a specified device; B6 verifying does
	not guarantee approval.
	\item[B8: Removal of sensitive data on decommissioning.] Means to delete, from
	the device, all sensitive data recorded by the device, plus all sensitive keying material
	acquired by the device from the domain.  
		\item[B9: Decommissioning support for transfer.] Means to support a new to-device access password, for low-risk transfer of IoT devices to second owners. 
\end{description}

\textbf{Device attestation.}
Device attestation (e.g., see \cite{rattanavipanon:2018:manuscript})
provides a means for the trust domain to test (part of) the device operating state, e.g., to verify the correctness of security-critical code, and detect unauthorized changes. While ideally desirable, especially as an advanced feature
in higher-value IoT devices 
in industrial environments,  
hardware support in less likely in commodity IoT devices near-term.
We thus consider device attestation 
beyond the scope of baseline security functionality for commodity 
IoT devices.
%


\section{Background and Related Work}  \label{sect:background}

We review IoT-relevant approaches for establishing and managing keying relationships with wireless devices.

\textbf{Ad-Hoc Key Agreement.}
Key sharing in ad-hoc settings has been a topic of active research for over 20 years.   In the \emph{resurrecting duckling} model \cite{Stajano:1999:workshop}, a device (duckling) is \emph{imprinted} by a master (mother), setting up a security association (shared \emph{ignition key}) when the device first starts up. The analogy is that a newly hatched duckling adopts (trusts) as its partent the first entity encountered. Such imprinting can be used by a device dedicated to managing a domain's IoT devices. To protect the imprinting from imposter management devices, a suggestion was physical electrical contact with the target device with the shared key transferred in plaintext over the electrical connection.
In parallel, Asokan 
\cite{Asokan2000} proposed a solution for ad-hoc key agreement 
for a multi-device scenario wherein a group of devices needs to establish a common key, based on an authentication password shared between the devices over an out-of-band channel (vs.\ 
device-to-device security associations \cite{Stajano:1999:workshop}).

\textbf{Industry-driven on-boarding efforts.}
BRSKI~\cite{I-D.ietf-anima-bootstrapping-keyinfra} and Intel Secure Device Onboard (Intel SDO) 
\cite{Brickell2007,Intel2017}  
are among efforts seeking a scalable approach (Section \ref{sect:requirements}) to key management by pre-installing cryptographic material on IoT devices. This supports mutual authentication between a device and a trust domain's registrar, and verification that the device is authorized to join the domain. The approach has additional requirements,
e.g., BRSKI pre-installs, into each IoT device, public-key based ``Pledge vouchers''. 
Intel's solution utilizes a dedicated hardware chip storing required key material and using remote attestation to assure authenticity and integrity of devices admitted to the trust domain.

Neither solution suits generic on-boarding of consumer/commodity IoT devices within our scope;
BRSKI explicitly declares its scope as non-constrained devices (above RFC 7228's Class 2 of Section~\ref{sect:iot-environments}).
Both pose relatively stringent requirements on hardware and organizational set-up between device vendors and trust domains, limiting application to specific IoT scenarios (e.g., smart city or industrial IoT per Section~\ref{sect:iot-environments}), where required key material can be mass-provisioned on devices pre-deployment, and administrative measures exist for associating and authenticating keys with individual domains. Such measures do not suit open IoT environments with thousands of distinct device manufacturers.

\textbf{Hardware security.}
Hardware-based security mechanisms, including \emph{trusted platform modules} (TPMs)
and \emph{trusted execution environments} (TEEs) such as  
ARM TrustZone \cite{guan:2017:mobisys,rattanavipanon:2018:manuscript},
%
can protect security-critical information on individual devices.
They offer two advantages:

\textit{1) Improved resilience to physical attacks.}
Hardware-rooted defenses raise the cost of extracting  from individual devices
both sensitive personal information 
and intellectual property (e.g., algorithms). 
However even thorough such mechanisms may fail to stop well-resourced adversaries with physical access to targeted devices.

\textit{2) Reduced size of Trusted Computing Base (TCB).}
By moving sensitive calculations to a TEE, processing of sensitive data is effectively decoupled from generic processing of applications on the device. Application software vulnerabilities then can't be used to exfiltrate sensitive data (e.g., keys) from the device, as all sensitive operations are encapsulated inside the TEE. Only results of computations with sensitive data (e.g., encrypted data) are returned outside the TEE. Thus the TEE provides protection against leaking of sensitive data
in the face of many software-based attacks.

An obstacle to relying on hardware-based security is that
in price-competitive commodity devices, even
small costs are barriers to wide adoption.
Protection against physical \emph{side-channel attacks} is also a concern, albeit
these often require targeted attack (and are thus less scalable). 
Design approaches that minimize the impact of key extraction are thus important---e.g., despite attempts to protect stored keys by hardware security measures, 
the risk in using global/master 
keys across devices was shown by the
Philips Hue smart lightbulb hack  \cite{Ronen2017}.

\textbf{Software Updates and Ownership.}
Wurster~\cite{wurster:2007:hotsec}  
discusses a 
mechanism for controlling updates to binaries. 
One or more public keys  
are embedded in key-locked objects; to confirm updates are from the same developer or publisher, signed updates must be verifiable by one or more keys in the installed version. 
Developer-generated key pairs are used on intial installations in a trust-on-first-use model, avoiding a centrally-managed PKI; the Android OS uses a version of this approach. 
Smartphone software installation and update \cite{Barrera:2011:SPmag}
offers lessons for the IoT world. 
%
 \textit{Meteor} \cite{barrera:2O12:most} is a
 mobile phone application installation framework
 for multi-market environments (in contrast to the
 single-market approaches of major smartphone vendors). 
\textit{Baton} \cite{barrera:2O14:wisec} offers \textit{certificate agility}
in a decentralized code-signing infrastructure, to address the requirement of
transferring signing authority recognized for software updates over time, 
an issue particularly relevant for long-lifetime IoT devices.
Khan \cite{Khan2017} explores automating
change of ownership of IoT devices;
 the \textit{SELIoT} project \cite{Seliot2017} explores IoT life-cycle security.

\textbf{Software Update Systems: TUF and successors.}
\textit{TUF} (The Update Framework),
and its successor \textit{Diplomat} \cite{kuppusamy:2016:nsdi}, aim to
provide resilience against compromised signing keys, while securing  
updates from software repositories 
and community repositories; 
the later \textit{Uptane} 
secure software update framework \cite{kuppusamy:2016a:escar} 
targets additional IoT-related complications specific to automotive sector updates.
A further update framework, \textit{CHAINIAC} \cite{nikitin:2017:usenix},  
offers additional protections by decentralizing various roles in TUF and Diplomat,
using more signing keys, facilitating agility and signing key evolution,
and providing transparency.  
IoT software/firmware update is the subject of an
Internet Architecture Board report \cite{tschofenig:rfc8240:iotsu}. 
\textit{ASSURED} \cite{rattanavipanon:2018:manuscript} is 
an IoT-focused firmware update framework designed to extend any update distribution
scheme (e.g., TUF), with additional functionality, e.g., verifying
successful update installation on target IoT devices.

\textbf{Efficient cryptography on constrained devices.}
Optimization of cryptographic primitives is a recurring theme, 
dating back to 8-bit smartcards in the 1980s. Recent studies,
including of elliptic-curve Diffie-Hellman (ECDH) on 8-bit microcontrollers \cite{Ozmen2017},
suggest that even constrained devices can support suitably-designed key agreement
protocols for on-boarding (Section \ref{sect:arch}).
Lightweight symmetric cryptography \cite{biryukov:2017:eprint}
is already suitably efficient for low-resource devices, 
but standardization efforts \cite{nist:lightweight:2017}
are only now emerging.







\section{Concluding Discussion} \label{sect:conclusion}


The suggested design for mediation of IoT-device software updates
significantly simplifies both the trust infrastructure needed in the short term,
and the management complexity of maintaining it over
long device lifetimes spanning corporate and technology evolution. 
This avoids a major issue 
still not fully addressed in the Internet of Computers (IoC):
management of a global PKI including trust anchors in end-devices,
and side issues like certificate revocation 
(still largely unsupported in TLS).  

The design also reduces in-device support needed for
public-key cryptography to that for an IoT-tailored PAKE
for master key establishment (B2 above),
and provides domain owners visibility and control over the devices within their
trust domains through the on-boarding process, and mediated software updates. 
To flexibly accommodate non-technical domain owners (e.g., home owners),
the \guardian model allows delegation of management to service
providers, allowing outsourcing.

Protection of IoT devices cannot depend solely on traffic filtering methods of IP traffic, 
as many devices have communication interfaces independent of the TCP/IP stack---e.g., 
the noted Philips Hue hack~\cite{Ronen2017} did not utilize the IP stack.
Risks to end-users increase with the number of consumer IoT devices with direct WAN access 
(independent of a local domain's wireless access). 

Completely removing all manufacturers from controlling software update is 
unrealistic. Premium manufacturers of IoT product families 
who focus on both usability and security, may insist on end-to-end control of software updates. 
Such ``Class A" companies,
with long-term plans and/or full-stack expertise, 
are also more likely to deploy devices with TEEs, to secure  
updates (including supporting crypto verification)
and intellectual property, e.g., 
algorithms that enhance battery life or communications; and 
to responsibly arrange updates and
manage trust anchors even should they vacate the IoT business.
Mediated updates do not suit them.  Thus our proposal targets 
``Class B" manufacturers competing mainly on price, with many
guaranteed to leave the IoT business before 10-20 year products 
expire, and lacking security expertise and motivation for long-term support.

Vendor-specific IoT device management solutions complementary
to baseline functionality proposed herein comes
at the cost of users running multiple management applications.
Thus 
standardization of IoT device interfaces and baseline functionality appears
essential to reduce IoT chaos in both usability and security. 

Our de-commissioning design with manufacturers (or a proxy) running 
centralized password databases raises questions: what 
parties are suitable to run on-boarding password service proxies for 
manufacturers; what incentives exist for ongoing product support
after an original IoT device manufacturer discontinues its business?

The design outlined herein aims not to promote a particular solution,
but as a strawman to raise issues requiring immediate attention,
to avoid problems 10-20 years out. We aim to (a) highlight 
difficult issues to be addressed, and (b) provide a thought-experiment on 
IoT device manufacturers who compete on price, 
and whose products will outlive the companies themselves.
%
Thus our focus on ``baseline requirements".
How do we, as computer and security techologists, help the public  
who buy commodity IoT products? Manufacturers will reject the idea of
government regulation, but have they any alternate solutions for commodity devices?

Given the poor security track record of 
commodity device manufacturers, a lack of responsibility for devices post-sale, and the 
certainty that many devices will outlive the business of their manufacturers, 
we suggest that baseline security requirements for commodity IoT devices require greater
attention from the security community, user communities worldwide, and government
regulators.

\appendix



%
%

\section{IoT Device Taxonomy}  \label{sect:taxonomy}  \label{sec:bigtable}



\subsection{IoT Device Characteristics}  \label{sect:properties}

As noted in Section \ref{sect:iot-environments}, here we summarize characteristics of typical IoT devices in smart home environments and smaller-scale industrial environments, as a main target of our key management solutions. Appendix \ref{sect:low-med-high} then uses these characteristics to group devices in three categories. 

\textbf{Communication Protocols.}
Many commodity IoT devices 
use WiFi to connect to the IP network via a local WiFi router. However, for devices operating without a permanent connection to the power grid, i.e., operating on battery power, WiFi may be too power-hungry for use over prolonged periods of time. Such devices resort to energy-optimised protocols like ZigBee, Z-Wave or Bluetooth Low Energy (LE), which  employ extensive duty-cycling and relatively low-throughput communications to preserve power.
For connectivity to IP-networks these devices may use a hub device (in the case of ZigBee and Z-Wave) or a smartphone (in the case of Bluetooth LE) to convey messages to other devices or applications.

Exceptions are battery-powered devices using WiFi (including smartphones) with relatively large batteries expected to be regularly recharged; and devices operated only intermittently, e.g., digital scales powered on only a few times daily, which can therefore remain in deep hibernation or completely powered off most of the time.

\textbf{Power.}
Many
IoT devices draw their power directly from the power grid. Such 
devices thus typically have limited mobility or are stationary.
Other devices, for which using a direct connection to the powergrid is not possible, due to special device placement must operate on battery power, e.g., if outdoors or in situations in which wiring would be cumbersome to install or aesthetically undesirable. 
Typical such devices include: door and window sensors, temperature and weather sensors, and light switches. These must operate for extended periods without requiring battery exchange and are thus typically quite limited in  functionality, e.g., dedicated to sensing and reporting a few contextual parameters to a hub device or smartphone.

\textbf{Encryption Capabilities.}
Virtually all devices using WiFi support the WPA2 encryption standard, enabling protection of wireless communications to the local WiFi access gateway.
Low-end devices typically lack computational capabilities to perform public-key operations and must therefore resort to symmetric ciphers to protect communications. Some recent ultra low-power MCUs offer hardware acceleration for strong cryptographic algorithms like AES-256, allowing even very low-end devices to employ strong encryption for communications. 
High-end devices often use public key certificates to authenticate to back-end services, and standard protocols like TLS to protect communications to the vendor site.

\textbf{Software Environment.}
High-end devices typically employ a full OS capable of multitasking and running programs and scripts executing arbitrary computations.
Low-end devices using microcontroller-based solutions are much more limited, typically directly executing firmware, with functionality limited to core device functions; introducing new functionality typically requires installing entirely new firmware.

\subsection{IoT Device Categories}  \label{sect:low-med-high}

As noted above, IoT Devices can be characterized according to characteristics such as  processing capability, memory and software 
platform.
From these factors, we distinguish  three rough categories of IoT devices---\emph{high-end}, \emph{mid-level}, \emph{low-end}. Table~\ref{tab:iot-devices} 
categorizes selected examples of consumer IoT devices and lists some device characteristics (e.g., communication capabilities, OS).

\textbf{High-End Devices.}
High end devices resemble for most parts general-purpose computers and are equipped with a CPU and sufficient memory to run an embedded multitasking-capable operating system. These devices can independently communicate with their respective vendor-provided back-end services and can perform relatively complex computation tasks. Typically they provide a vendor-specific set of services with a related smartphone companion app. Administrative access to the device and services are provided through the companion app or a web-based interface served by the back-end service. Examples of high-end devices indlude home assistants and entertainment hubs like Google Home, Amazon Alexa and Apple TV.

\textbf{Mid-Level Devices.}
Mid-level devices are equipped with a CPU and an OS (e.g., an embedded version of Linux), but have in general more limited processing power and memory available. Their functionality is therefore limited to the actual functionality of the device. They may serve a generic (web-based) administration interface to the user for administrative access (e.g., many IP cameras are administered this way) or, administrative access may be facilitated with the help of a companion app. Examples of mid-level devices are shown in Tab.~\ref{tab:iot-device-details}. The D-Link DCS-930L IP camera, Edimax SP-2101W smart power plug, Philips Hue lighting bridge and D-Link DCH-S150 WiFi Motion sensor are typically running a Linux-based operating system and have a relatively powerful CPU with a moderate amount of memory that can be used to perform various general-purpose computations.

\textbf{Low-End Devices.}
These are commonly based on microcontroller-driven SoC designs. This typically results in limited processing power and insufficient memory to host a general-purpose OS, limiting functionality to core device functions. The Smarter iKettle 2.0 water kettle and Netatmo Weather station are MCU-based designs supported by dedicated WiFi chips---see Table~\ref{tab:iot-device-details} for example mid-level and low-end devices.


\balance  


\renewcommand{\bottomfraction}{0.9}
\renewcommand{\textfraction}{0.1}

\begin{table*}[!bpt] 
	\centering
			\caption{Hardware/software details for selected mid-level and low-end IoT devices. The aim is to provide general context.} 
	\label{tab:iot-device-details}
	\vspace{12pt}
	\begin{threeparttable}
		\begin{tabular}{b{4cm}b{5cm}b{4cm}l}
		Device Model                                           & CPU        & Memory & OS \\ 
		\hline
		\hline
		\textbf{Mid-level devices} \\ \hline5
		D-Link WiFi Day Camera DCS-930L\tnote{a}                        
																& Ralink RT5350F CPU 360 MHz & 32 MB RAM 4 MB Flash     & Linux 2.6.21 \\ \hline
		Edimax SP-2101W Smart Plug Switch\tnote{b} 
																& Ralink RT5350F CPU 360 MHz & 32 / 16 kB cache, up to 64 MB SDRAM, 32 Mbit serial Flash                       & Linux 2.6.21 \\ \hline
		Philips Hue Bridge model 3241312018\tnote{c}                   
																& Qualcomm QCA4531 CPU 650 MHz & 64MB RAM, 1Gb flash & OpenWRT Linux \\
																& Atmel ATSAMR21E IEEE 802.15.4 (ZigBee) Cortex-M0+ 48MHz & max. 32 kB SRAM max. 768 kB Flash
																															 \\ \hline
																									
		D-Link WiFi Motion sensor DCH-S150                     & WiFi SoC 650MHz            & 32MB DDR SRAM                   &  OS unknown\\ \hline
								 \textbf{Low-end devices}\\ \hline
		Smarter iKettle 2.0 water kettle SMK20-EU              & AI-Thinker ESP82266MOD WiFi+MCU 80MHz MCU & 64kB ROM 32 kB RAM 32kB cache 80kB user data RAM & - \\
																& STM32F015C8T6 48MHz MCU  & 16-64 kB Flash 8kB SRAM      &  \\ \hline
		Netatmo weather station                                & STM32F103 MCU ARM 32b-Cortex-M3 72MHz & 64-128 kB Flash 20 kB SRAM & - \\
		& USI WM-BN-BM-04 WiFi + BTLE + FM & & \\ \hline
	D-Link DCH-Z110 Door Sensor & SD3502 General Purpose SoC Integrated MCU and RF tranceiver & 16 kB SRAM 128 kB Flash & - \\ \hline
	\end{tabular}
	\begin{tablenotes}
		\begin{footnotesize}
			\item [a] \url{https://wiki.openwrt.org/toh/d-link/dcs-930l}
		\item [b] \url{https://community.openenergymonitor.org/t/teardown-of-edimax-smart-plug-sp-2101w-us-version/2098}
		\item [c] \url{https://www.reddit.com/r/huelights/comments/3nsx4b/hue_hub_v20_teardown/}
		\end{footnotesize}
	\end{tablenotes}
	\end{threeparttable}
\end{table*}


\newcommand{\mcrot}[4]{\multicolumn{#1}{#2}{\rlap{\rotatebox{#3}{#4}~}}} 
\newcommand{\bul}{\textbullet}
\newcommand{\cir}{$\circ$}

\begin{table*}
	\centering
	\caption{Classification of IoT devices (high-end, mid-level, low-end).
	Extremely-constrained devices (per RFC 7228) are below low-end. Commodity devices within our scope are  
	typically mid-level or low-end, but may include high-end devices not managed by premium manufacturer solutions.} 
	\label{tab:iot-devices}
	
	\begin{threeparttable}
		\begin{tabular}[!th]{lllcc|cc}
		Identifier         & Device Model                                           & \mcrot{1}{l}{60}{WiFi} & \mcrot{1}{l}{60}{Ethernet} & \mcrot{1}{l}{60}{LE protocol(s)\tnote{a}} & \mcrot{1}{l}{60}{OS\tnote{b}}     & \mcrot{1}{l}{60}{Battery-operated} \\
		\textbf{high-end}  &  \\ \hline
		AmazonEcho         & Amazon Echo                                            & \bul                   & \cir                       & \cir                             & \bul                     & \cir                         \\
		AmazonEchoDot      & Amazon Echo Dot                                        & \bul                   & \cir                       & \bul                             & \bul                     & \cir                         \\
		ApexisCam          & Apexis IP Camera APM-J011                              & \bul                   & \bul                       & \cir                             & \bul                     & \cir                         \\
		AppleTV            & Apple TV                                               & \bul                   & \bul                       & \bul                             & \bul                     & \cir                         \\
		CamHi              & Cooau Megapixel IP Camera                              & \bul                   & \bul                       & \cir                             & \bul                     & \cir                         \\
		D-LinkCamDCH935L   & D-Link HD IP Camera DCH-935L                           & \bul                   & \cir                       & \cir                             & \bul                     & \cir                         \\
		D-LinkCamDCS930L   & D-Link WiFi Day Camera DCS-930L                        & \bul                   & \bul                       & \cir                             & \bul                     & \cir                         \\
		D-LinkCamDCS932L   & D-Link WiFi Camera DCS-932L                            & \bul                   & \bul                       & \cir                             & \bul                     & \cir                         \\
		EdimaxCamIC3115    & Edimax IC-3115W Smart HD WiFi Network Camera           & \bul                   & \bul                       & \cir                             & \bul                     & \cir                         \\
		EdnetCam           & Ednet Wireless indoor IP camera Cube                   & \bul                   & \bul                       & \cir                             & \bul                     & \cir                         \\
		GoogleHome         & Google Home                                            & \bul                   & \cir                       & \cir                             & \bul                     & \cir                         \\
		SmcRouter          & SMC router SMCWBR14S-N4 EU                             & \bul                   & \bul                       & \cir                             & \bul                     & \cir                         \\
		UbnTAirRouter      & Ubnt airRouter HP                                      & \bul                   & \bul                       & \cir                             & \bul                     & \cir                         \\
		WansviewCam        & Wansview 720p HD Wireless IP Camera K2                 & \bul                   & \cir                       & \cir                             & \bul                     & \cir                         \\ \hline
		\textbf{mid-level} &  \\ \hline
		D-LinkSensor       & D-Link WiFi Motion sensor DCH-S150                     & \bul                   & \cir                       & \cir                             & \bul                     & \cir                         \\
		D-LinkSiren        & D-Link Siren DCH-S220                                  & \bul                   & \cir                       & \cir                             & \bul                     & \cir                         \\
		D-LinkSwitch       & D-Link Smart plug DSP-W215                             & \bul                   & \cir                       & \cir                             & \bul                     & \cir                         \\
		D-LinkWaterSensor  & D-Link Water sensor DCH-S160                           & \bul                   & \cir                       & \cir                             & \bul                     & \cir                         \\
		EdimaxPlug1101W    & Edimax SP-1101W Smart Plug Switch                      & \bul                   & \cir                       & \cir                             & \bul                     & \cir                         \\
		EdimaxPlug2101W    & Edimax SP-2101W Smart Plug Switch                      & \bul                   & \cir                       & \cir                             & \bul                     & \cir                         \\
		D-LinkHomeHub      & D-Link Connected Home Hub DCH-G020                     & \bul                   & \bul                       & \bul                             & \bul                     & \cir                         \\
		HueBridge          & Philips Hue Bridge model 3241312018                    & \cir                   & \bul                       & \bul                             & \bul                     & \cir                         \\
		iKettle2           & Smarter iKettle 2.0 water kettle SMK20-EU              & \bul                   & \cir                       & \cir                             & \cir                     & \cir                         \\
		Lightify Gateway   & Osram Lightify Gateway                                 & \bul                   & \cir                       & \bul                             & \cir                     & \cir                         \\
		Netatmo            & Netatmo weather station                                & \bul                   & \cir                       & \bul                             & \cir                     & \cir                         \\
		SmarterCoffee      & Smarter SmarterCoffee coffee machine SMC10-EU          & \bul                   & \cir                       & \cir                             & \cir                     & \cir                         \\
		TP-LinkPlugHS100   & TP-Link WiFi Smart plug HS100                          & \bul                   & \cir                       & \cir                             & \bul                     & \cir                         \\
		TP-LinkPlugHS110   & TP-Link WiFi Smart plug HS110                          & \bul                   & \cir                       & \cir                             & \bul                     & \cir                         \\
		WeMoInsightSwitch  & WeMo Insight Switch model F7C029de                     & \bul                   & \cir                       & \cir                             & \bul                     & \cir                         \\
		WeMoLink           & WeMo Link Lighting Bridge model F7C031vf               & \bul                   & \cir                       & \bul                             & \cir                     & \cir                         \\
		WeMoSwitch         & WeMo Switch model F7C027de                             & \bul                   & \cir                       & \cir                             & \bul                     & \cir                         \\
		Withings           & Withings Wireless Scale WS-30                          & \bul                   & \cir                       & \cir                             & \cir                     & \cir                         \\
		Aria               & Fitbit Aria WiFi-enabled scale                         & \bul                   & \cir                       & \cir                             & \cir                     & \cir                         \\ \hline
		\textbf{low-end}   &  \\ \hline
		D-LinkDoorSensor   & D-Link Door \& Window sensor                           & \cir                   & \cir                       & \bul                             & \cir                     & \bul                         \\
		EveRoom            & Eve room sensor                                        & \cir                   & \cir                       & \bul                             & \cir                     & \bul                         \\
		EveWeather         & Eve weather sensor                                     & \cir                   & \cir                       & \bul                             & \cir                     & \bul                         \\
		FibaroMotionSensor & Fibaro motion sensor                                   & \cir                   & \cir                       & \bul                             & \cir                     & \bul                         \\
		HomeMaticPlug      & Homematic pluggable switch HMIP-PS                     & \cir                   & \cir                       & \bul                             & \cir                     & \cir                         \\
		HueSwitch          & Philips Hue Light Switch PTM 215Z                      & \cir                   & \cir                       & \bul                             & \cir                     & \bul                         \\
		Osram Lightify     & Osram smart light bulb                                 & \cir                   & \cir                       & \bul                             & \cir                     & \cir                         \\
		NetatmoWind        & Netatmo wind gauge                                     & \cir                   & \cir                       & \bul                             & \cir                    & \bul                         	\end{tabular}
\begin{tablenotes}
	\item [a] \bul = supports one or more low-energy communication protocols
	\item [b] \bul = OS resident, \cir = no OS provided
\end{tablenotes}
\end{threeparttable}
\end{table*} 

\end{document}